# Characterization of $Al_{12}Mg_{17}$ Nanofluid by Dynamic Light Scattering and Beam Displacement Methods

**Soroush Javadipour[1], Ali Shokuhfar[1], Zeinab Heidary[2], Mohammad Amin Amiri Roshkhar [3], Keyvan Homayouni[4], Fatemeh Rezaei[5], Ashkan Zolriasatein[6], Shahrokh Shahhosseini[7], Alimorad Rashidi[8], S. M. Mahdi Khamoushi[3]**

[1] Faculty of Materials Science and Engineering, K. N. Toosi University of Technology, 15 Pardis St., Tehran 1991943344, Iran.

[2] Department of Mechanical Engineering, K. N. Toosi University of Technology, 15 Pardis St., Tehran, 1999143344, Iran.

[3] Laser and Plasma Research Institute, Shahid Beheshti University, Tehran, Iran.

[4] Department of Petroleum Engineering, Islamic Azad University, Science & Research Branch, Tehran, Iran.

[5] Department of Physics, K. N. Toosi University of Technology, 15875-4416 Tehran, Iran.

[6] Non-metallic Materials Research Department, Niroo Research Institute, Tehran, Iran.

[7] Department of Chemical Engineering, Iran University of Science and Technology, Tehran, Iran.

[8] Department of Chemical Engineering, Research Institute of Petroleum Industry, Tehran, Iran.

*Corresponding authors*: [s.javadipour2020@gmail.com](mailto:s.javadipour2020@gmail.com), [fatemehrezaei@kntu.ac.ir](mailto:fatemehrezaei@kntu.ac.ir)

## Abstract

The thermal conductivity and stability of nanofluids have posed the biggest challenges to their adoption as coolants in thermal applications in industries such as electronic equipment, heat exchangers, and solar technologies. In this paper, the thermal conductivity coefficient of an $Al_{12}Mg_{17}$ nanofluid is measured by a novel beam displacement method. Besides, the stability, particle size distribution (PSD), TEM micrograph, and electrical conductivity of $Al_{12}Mg_{17}$ nanofluids are investigated. For the preparation of nanofluids, three different surfactants are used to disperse $Al_{12}Mg_{17}$ nanoparticles in DI water using two-step method. Then, dispersion stability is monitored visually and quantified using a zeta potential test. The thermal conductivity coefficient and particle size distribution are measured using two optical setups. For the purpose of evaluating the outcomes, the thermal conductivity coefficients estimated using the beam displacement method are compared with the KD2 Pro apparatus results, and the PSD findings are examined using TEM micrographs. Results demonstrate that a 1:1 ratio of CTAB and $Al_{12}Mg_{17}$ nanoparticles is proper for stabilizing $Al_{12}Mg_{17}$ nanofluid. Also, the optimum ultrasonication period is determined to be 2 hours, and the peak of particle size distribution is measured to be 154 nanometers at this time. Thermal conductivity measurements show that the thermal conductivity coefficients improve as the concentration of $Al_{12}Mg_{17}$ nanoparticles increases, reaching a maximum enhancement of 40% in comparison to the base fluid at a concentration of 0.05 vol.%. Also, the electrical conductivity outcomes indicate a linear increase from 155 to 188 $\mu S/cm$ by enhancing the nanoparticles concentration from 0.0125 to 0.05 vol.%. Among the various techniques, optical methods for determining the thermal conductivity coefficient of nanofluids have the advantage of giving results before the development of bulk motion. Therefore, the administration of nanofluids in thermal applications requires the development of optical techniques to provide more accurate measurements.

## Introduction

A nanofluid is a heterogeneous mixture of a base fluid and nanoparticles which can be utilized in a broad range of thermal applications in the industry [1] and medicine [2], such as solar collectors [3], vehicle radiators [4], electronic cooling [5]. In many of these applications, nanofluids play a substantial and vital role in the system performance because of their efficiency at transferring heat, so the attention of engineers and scientists have been attracted in studying of the heat transfer into the systems. The differences in thermal conductivity between nanofluids have been studied in many works of literature [6]. Therefore, it is necessary to characterize the different properties of nanofluids including thermal and electrical properties, stability, and particle size distribution for applying in industry.

Regarding thermal characterization of nanofluids, the scientists have used different methods for determination of the thermal conductivity such as the transient methods, three omega [7], temperature oscillation [8], steady state parallel plate [9], thermal comparator [10], and optical methods which each of them have a difference in method of determination. For example, the transient hot-wire (THW) [11] and transient plane source (TPS) [12] are based on monitoring the temperature of heat source and time response after being exposed into an electrical pulse [13]. Furthermore, thermocouples are used in the steady state methods, and it is important to keep the temperature reading discrepancies to a minimum when the thermocouples are at the same temperature [10]. In addition, in thermal comparator, evaluation of the sample's conductivity needs only one point of contact [10]. Another approach in order to determine thermal conductivity is the optical methods which are based on the interaction between light and fluid.

Generally, a number of optical methods, such as the laser flash analysis (LFA) technique [14], are used to measure the thermal conductivity of nanofluids. Moreover, there are other optical methods including beam deflection methods [15] and hot-wire laser beam displacement techniques that rely on the temperature-dependent deflection angle in nanofluids [16]. The hot-wire laser beam displacement method can assess the thermal conductivity and thermal diffusivity of nanofluids [16]. Generally, beam displacement method is based on changing the reflecting index by temperature variations so that the thermal conductivity and thermal diffusivity of nanofluids increase with an increase in volume fraction [16, 17].

Several research groups have reported the variations of the thermal conductivity of different nanofluids, versus the type, size, and the base fluid. Furthermore, different researchers have employed a variety of nanoparticles, including single-element, single-element oxide, multi-element oxides, metal carbides, metal nitrides, and carbon base nanoparticles. These nanoparticles can be dissolved in various liquids such as water, ethanol, EG, oil, and refrigerants [18-20]. For instance, Paul et al. [20] mechanically alloyed $Al_{95}Zn_{05}$ before using a two-step process to disperse the nanoparticle in ethylene glycol. According to the thermal conductivity characterization, 0.10 vol% dispersion of nanoparticles resulted in a 16% increase in thermal conductivity relative to the base fluid. In our last work, the thermal and electrical conductivity of $Al_2O_3$-ZnO-CNT nanoparticles dispersed in DI water was explored [21]. Also, the particle size distribution of $Al_2O_3$-ZnO nanofluid versus time was investigated in order to study the stability of the nanofluid. It was concluded that by adding carbon nanotubes to $Al_2O_3$-ZnO nanofluid and forming 0.05 wt.% hybrid nanofluid, the thermal conductivity coefficient was enhanced by 30% in comparison with DI water. In addition, An experimental study on the thermal conductivity of $Cu_5Zn_8$ nanoparticles dispersed in an oil-based fluid was conducted by Farbod et al. [22]. Their result demonstrated that, in comparison to the base fluid, oil-based nanofluids with various concentrations of $Cu_5Zn_8$ nanoparticles had better thermal conductivity. It should be mentioned that little intermetallic materials have been employed by researchers as nanofluids, like Ti-6Al-4V [23], $Al_{95}Zn_{05}$ [20], $Cu_5Zn_8$ [22], and NiAl intermetallic nano-powders [24]. In addition, $Al_{12}Mg_{17}$ is another intermetallic substance which has been the subject of a few studies [25].

An essential issue for researcher in the evaluation of nanofluid performance is the relationship between ultrasonication period length and the size of the nanoparticles in the nanofluid. Sonication is a physical technique that can be applied to improve the stability of the nanofluid by rupturing the attractional force of the nanoparticles in order to reduce their size [26]. Dynamic light scattering is one of the most accessible and practical methods for estimating the sizes and their distributions [27]. For instance, Poli et al. [28] studied the relationship between the particle size of SAz-1 and SWy-1 montmorillonite and the ultrasonication period. According to their results, after 60 minutes of sonication, SAz-1 exhibited a relatively broad particle size distribution with a modal diameter of 283 nm. Additionally, after being sonicated for 60 minutes, the SWy-1 displayed a bimodal distribution of particles at 140 and 454 nm. Using dynamic light scattering, Afzal et al. [29] reviewed the impact of ultrasonication period on the average particle size of various nanofluids. According to their observations, lengthening the sonication process reduces the particle size, improves dispersion, and increases stability. Furthermore, the optimum ultrasonication period was accomplished, resulting in the highest performance. It has been discovered that ultrasonication periods longer than the ideal one do not improve the stability.

According to the literature review, a thorough investigation of the physical and thermal properties of $Al_{12}Mg_{17}$ nanofluid has only been done in a limited number of researches. In thermal applications involving nanofluids, the performance of the heat transfer procedure is dependent on the stability of the nanofluid as well as its thermal properties. Therefore, in this study, in addition to the physical properties of $Al_{12}Mg_{17}$ nanofluid including stability, optimum ultrasonication period, and electrical conductivity, the thermal conductivity is assessed using an innovative beam displacement method. This method has the novelty of focusing on the complex effect of nanofluid's thermal properties on the light beam deviation using an image processing technique. As part of the novelty of the present research, CCD detectors are used instead of PSD detectors for beam displacement detection, since virtual methods may not be as reliable as intuitive methods. The nanofluids are prepared in the current study utilizing a two-step method, and the zeta potential test and sediment observation are then employed to consider the effect of variation in type and weight percentage of surfactant on the stability of the $Al_{12}Mg_{17}$ nanofluids. Nanofluids exhibit complicated time-dependent behavior after ultrasonication which can be attributed to their size and morphology. There is also a lack of research that evaluates the variations in nanoparticle size distribution versus the ultrasonication period. In this regard, the dynamic light scattering (DLS) method is exploited to analyze the particle size distribution of the nanofluid versus the ultrasonication period. Moreover, transmission electron microscopy (TEM) is employed to examine the size and morphology of the $Al_{12}Mg_{17}$ nanoparticles and the validity of the DLS findings. Besides, the thermal conductivity outcomes from the beam displacement method are compared with the nanofluid's thermal conductivity coefficients measured by the KD2 Pro apparatus. It is shown that the $Al_{12}Mg_{17}$ nanoparticles in DI water as the base fluid, stabilized by the surfactant and ultrasonication, significantly affect the thermal properties of the nanofluid.

## Materials and methods

### Preparation method

In this research, the $Al_{12}Mg_{17}$ nanoparticles are employed which have an average particle size of 24 nm after 20 hours of milling. Additional information about physical properties of $Al_{12}Mg_{17}$ is available in references [25, 30]. A two-step process was used to disperse the nanoparticles in the base fluid in order to create $Al_{12}Mg_{17}$ nanofluids. The two-step procedure involves producing nanoparticles separately first, and then, dispersing the created nanoparticles in the base fluid using a variety of physical treatment methods, such as utilizing a stirrer, an ultrasonic bath, and an ultrasonic disruptor [31].

In this study, the $Al_{12}Mg_{17}$ nanofluids with varying volume concentrations were created by diluting the concentrated suspensions. Table 1 summarizes the details of the preparation procedure. Here, three surfactants, PVA (polyvinyl alcohol), SDS (sodium dodecyl sulfate), and CTAB (cetyltrimethylammonium bromide) were used to produce the nanofluids. Moreover, five different concentrations of CTAB surfactant including 0.02, 0.04, 0.06, 0.08, and 0.1 wt.% were added into 100 ml of DI water.

Table. 1. Details of the methods utilized for producing nanofluids by two-step technique.

| Preparation process | Process details |
|---|---|
| Stirrer | Revolution speed: 900 rpm<br>Revolution time: 15 min |
| Ultrasonic disruptor | Sonication time: 30 min<br>Frequency: 60 kHz<br>Max. sonicating power: 250 W |
| Ultrasonic bath | Sonication time: 30 min<br>Frequency: 40 kHz<br>Max. sonicating power: 150 W |

### Testing procedure

First, 0.1 vol.% of PVA, SDS, and CTAB were individually dissolved in DI water. Then, 0.05 vol.% of $Al_{12}Mg_{17}$ nanoparticles were added to each solvent during its stirring. Under the conditions described in Table 1, both of the ultrasonic disruptor and ultrasonic bath were employed to disperse nanoparticle clusters into the nanofluids. When using the observational approach to assess the stability of nanofluids containing various surfactants, the nanofluid including CTAB showed appropriate stability. As a result, CTAB was used as the surfactant in all subsequent tests, and its optimal amount was chosen by comparing the zeta potential values at various concentrations. After selecting the optimal surfactant concentration, the optimum ultrasonication period was determined by utilizing the DLS technique to follow the particle size distribution. The DLS results in the current study were validated using TEM microscopy images. Final step of the test protocol involved measuring of the thermal and electrical conductivities for different mass fractions of $Al_{12}Mg_{17}$ nanofluids. Fig. 1 depicts the procedure's flowchart.

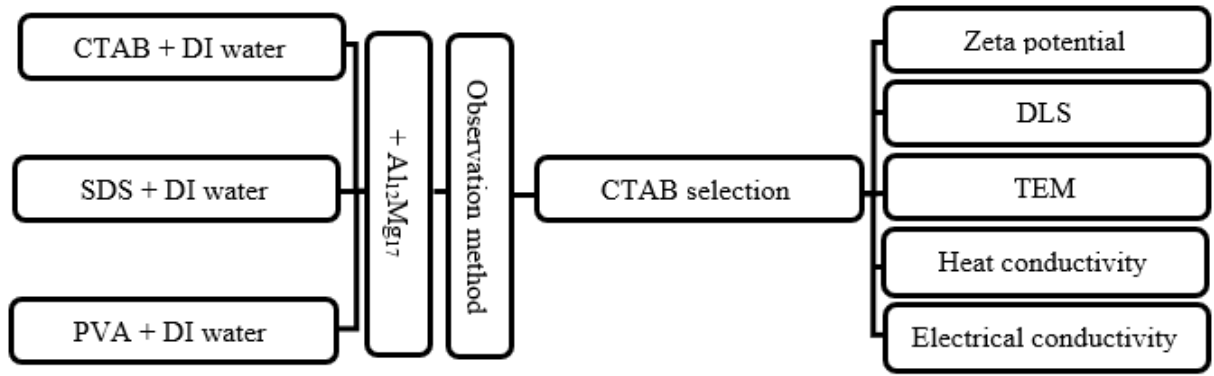

Figure 1. Different steps of the testing procedure: preparation and characterization of the nanofluid.

**Stability measurements**
Various techniques have been used by researchers to evaluate the stability of the nanofluids, including; observational methods [32], measurements of UV light absorbance spectra [33], pH values [34], and zeta potential analysis [35]. In this study, the stability of the $Al_{12}Mg_{17}$ nanofluid is assessed utilizing the observation method, and the zeta potential test. An observation method is based on taking pictures of a sample which contains generated nanofluid in a glass tube as time passes. It should be mentioned that it is possible to compare the collected images together and follow the sedimentation rate of the nanofluid [36]. Zeta potential testing is another method for evaluating the stability of nanofluids which involves measuring each sample of the $Al_{12}Mg_{17}$ nanofluid's zeta potential using a ZETA-check apparatus (Particle Metrix, Germany). Zeta potential can be calculated using the Henry equation by employing the electrophoretic mobility of particles in suspension [37]. In this research, the reported zeta potential values are the averages magnitudes obtained from three replicates of each sample.

**Particle size distribution measurements**
The dynamic light scattering method is a non-imaging method for figuring out the average size and the size distribution of particles in suspension or polymers in solution. The size of particles between one nanometer and several micrometers can be determined using this method. A coherent light source, such as a laser will disperse in diverse directions when it strikes a solution medium that contains particles. This kind of scattering for nanoparticles is called Rayleigh scattering, since the diameter of nanoparticles is significantly smaller than the wavelength of the laser (i.e, 632.8 nm), while for larger particles it follows the patterns of Mie scattering and Fraunhofer diffraction, respectively. The intensity of the scattered light fluctuates over time as a result of the Brownian motion of the particles inside of the solution. It should be noted that the frequency of variations in the intensity of scattered light by the larger particles is lower because of their slower rate of movement in the solution, and vice versa. A typical graph is produced by computing the autocorrelation function (ACF) of the intensity, causing its decay curve depend on the particle size. Furthermore, the size and morphology of the $Al_{12}Mg_{17}$ nanoparticles in the nanofluid are examined, and as well as the validity of the DLS findings are determined using transmission electron microscopy (TEM, Zeiss EM-10C).

## Theory of the dynamic light scattering method

The DLS technique involves the employment of a coherent laser beam to excite a sample including Brownian-moving particles in a solution. Particles in the path of the laser beam scatter light in different directions. The detector records the variations in the light intensity scattered over time that occur at a specific angle with respect to the incident beam propagation direction. The light scattered from moving particles can provide information about their movement patterns. The diffusion coefficient increases as the frequency of fluctuations in the scattering intensity enhance, and vice versa.

The relationship between the particle size and its thermal movement pattern is the foundation of the dynamic light scattering approach. The Stokes-Einstein equation presents the definition of this relationship [38] as:

$$D = \frac{K_B T}{3\pi \eta d} \tag{1}$$

here, $D$ is the diffusion coefficient, $d$ is the particle's hydrodynamic diameter, $\eta$ is the viscosity of the solution, and $k_B T$ is thermal energy. This equation expresses that a smaller particle size ($d$) causes a faster movement in the solution and a larger diffusion coefficient and vice versa.

In addition, calculating the autocorrelation function (ACF) of the detector's signal is necessary to derive the quantitative relationship between the intensity and diffusion coefficient. This can be typically accomplished with the correlator device. The ACF fitting function displays an exponential decaying relation as Eq. (2) [38]:

$$G(\tau) = \exp(-\Gamma \tau) \tag{2}$$

where, $\tau$ is decay time, $G(\tau)$ is the ACF function, and $\Gamma$ is the decay rate of this function. The decay rate and the diffusion coefficient of the sample particles in the solution are related below [38]:

$$D = \frac{q^2}{\Gamma} \tag{3}$$

where, $q$ is the scattering vector which is defined as [38]:

$$q = \frac{4\pi n_0}{\lambda_0} \sin(\frac{\theta}{2}) \tag{4}$$

where, $\lambda_0$ is the laser's wavelength, $n_0$ is the solution's refractive index, and $\theta$ is the observation angle of scattering. The diffusion coefficient can be obtained by replacing $q$ from

Eq. (4) in Eq. (3), and the hydrodynamic diameter of the sample particles can be calculated from Eq. (1).

Experimental Set-up for PSD measurements

In this paper, nanoparticles are excited using a 632.8 nm He-Ne laser as shown in the schematic of the experimental set-up in Fig. 2. The laser beam passes through a vertical polarization after passing through a linear polarizer. After traveling through the second polarizer for intensity adjustment, the laser beam is focused by the lens on the sample cuvette containing nanoparticles suspended in solution. Another lens with a 75 mm focal distance is used to focus the scattered light at an angle of 90 degrees with respect to the direction of beam propagation. A PMT detector with a gain of 10,000 records the fluctuations of the collected light intensity. An aperture with a diameter of 1 mm is positioned in front of the collector lens to ensure the formation of an appropriate coherence region on the detector surface. Finally, the signal received from the PMT after pre-amplifying is retrieved with the aid of a digital oscilloscope at a frequency of 100 kHz and then, forwarded into the computer to perform various studies such as computing the autocorrelation function (ACF).

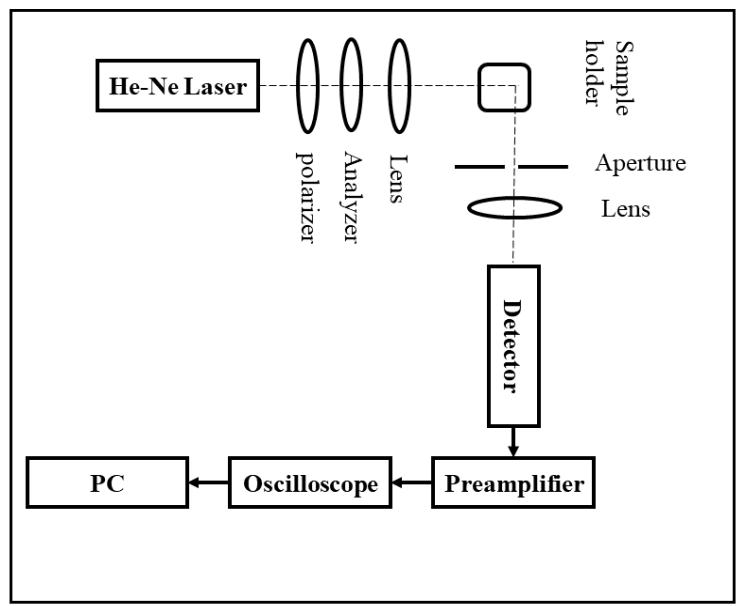

Figure 2. A schematic view of the DLS set-up utilized for PSD measurements.

Thermal conductivity measurements

In this research, the determination of thermal conductivity is performed by the usage of a novel beam displacement technique. Here, by applying a heat source to a nanofluid, the temperature changes affect the fluid's density and, as a result, its refractive index. When the beam travels through a path near the cylindrical heater, these variations lead to a beam deviation. The beam displacement method presented in the current research has three main steps: experimental evaluation of the beam displacement, numerical analysis of the temperature variations in nanofluid, and thermal conductivity calculation. In the first step, an optical set-up is designed to record the beam displacement after the deviation of the beam due to applying a thermal shock into the nanofluid. In the second step, a numerical simulation is conducted to calculate the temperature variations through a line corresponding to the beam path. In the third step, a trial-and-error loop is utilized to compare the experimental results with the numerical simulations which obtain the thermal conductivity of nanofluid. More details for each step are provided in the following sections.

Experimental evaluation of the beam displacement
Fig. 3 depicts the experimental layout used for the beam displacement measurements. A He-Ne laser with a 2-mW output at a wavelength of 632 nm is employed in the experiments. The polarizer and analyzer regulate the intensity of the laser light. A lens with a 50 mm focal length concentrates the laser beam on the samples. Additionally, two pinholes with hole widths of 50 and 20 microns are placed respectively for guiding the laser beam, and alignment is accomplished by the XYZ-stage, and the heater is secured by a specifically built cap on the cuvette. A CCD detector detects the beam displacement, and the outputs are sent to the PC. As part of the novelty of the present research, since virtual methods may not be as reliable as intuitive methods, CCD detectors are used instead of PSD detectors.

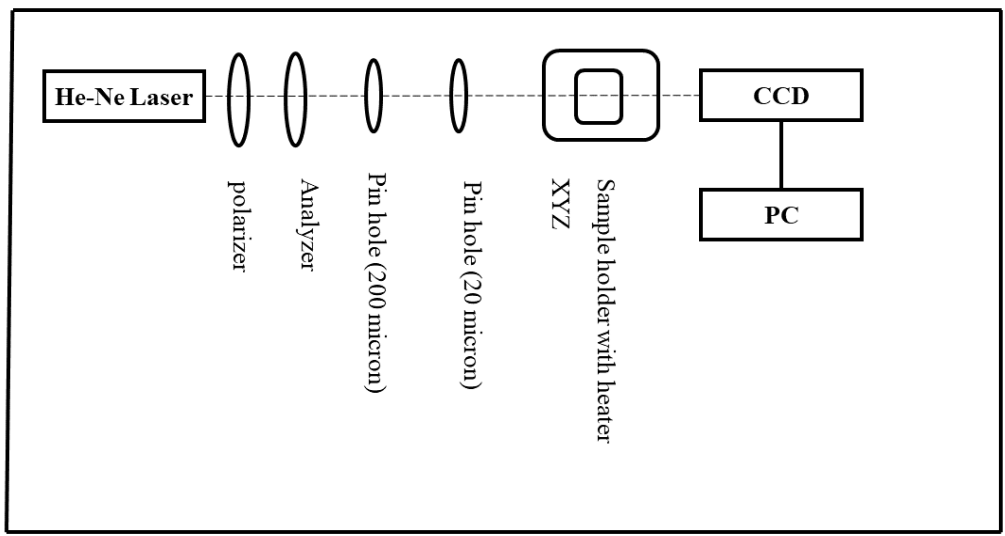

Figure 3. A Schematic of the experimental set-up designed for beam displacement measurement.

In order to evaluate the thermal conductivity magnitudes, Ali et al. [16] proposed the heat transport equations relevant to the beam displacement method as below:

$$\delta_T = \frac{w - W + n_0 L_{air}}{n_0} \frac{dn}{dT} \int_{-W}^{W} \frac{D}{\sqrt{D^2 + Z^2}} \frac{dT}{d\rho} dz. \tag{5}$$

In this equation, $\delta_T$ is the beam displacement, $w$ is the half width of the cuvette, $W$ is the inner size of the cuvette within which the probe beam deflected, $L_{air}$ is the distance between the cuvette and CCD detector, $n$ is the nanofluid's refractive index which depends on temperature, $n_0$ is the normal value, $\rho$ is the radial coordinate, $z$ is the spatial direction parallel to the original beam path, $D$ is the distance between the center of heater and probe beam, $Z$ is the adjacent side of a triangle with $\rho$ as hypotenuse and $D$ as opposite side, and $dT/d\rho$ is the temperature distribution which is obtained by a numerical simulation using finite element method (FEM). In addition, $\delta_T$ is extracted through an image processing method. The CCD detector is employed for the image processing, and the location of the brightest spot is found by Python software. It should be noted that the initial location of the brightest spot on the screen is recorded before the thermal shock was administered. After 5 seconds of exposure into the thermal shock, the beam deflection is recorded by measuring the displacement of the brightest spot. The fluid surrounding the heater's surface will experience greater temperature changes, hence, the beam must travel only a very small distance from the heater. In the current experiment, the heater's surface is 100 microns away from the place where the beam passed. Furthermore, controlling the light intensity is another crucial factor. If fewer beams pass close to the heater, the beam displacement is easier to be detected. Here, a CMOS sensor captures the beam spot intensity, and the Python OpenCV package is then applied to analyze the data. The sampling frames came into the computational process to measure the beam displacement during experiments. Moreover, the intensity profile is fitted using a Gaussian function to increase measurement accuracy with a resolution of 100 nanometers.

Numerical analysis of the temperature variations by FEM

The second step of the beam displacement method is evaluation by simulating the thermal variations on the beam path line. The computational study of the temperature evolutions in the numerical model is performed using finite element method. As depicted in Fig. 4, a 3D geometry of the cuvette containing fluid and cylindrical heater was modeled in accordance with the actual dimensions of the experimental set-up. To define the material properties, the cuvette and cylindrical heater were assumed to be solid bodies with specified densities, thermal conductivities, and heat capacities (Table 2). Corresponding material constants for the fluid are presented in Table 2 and were employed for simulating the fluid's thermal behavior in the present study. To ascertain the fluid's thermal conductivity, it was required to run simulations for a variety of thermal conductivity values in order to compare the results with the outputs of the beam displacement experiments.

In this paper, the thermal insulation conditions were applied for the exterior boundaries and the heater body is assumed to be a heat source with a heat generation power of 90 $MW/m^3$. As shown in Fig. 4, to determine the temperature variations, a 150 microns line from the heater surface was defined to correspond to the trajectory of the beam in the tests. All components were assumed to have a starting temperature of 298 K, which matches the temperature of the experiment's environment. Furthermore, the heat transfer equation was applied to the physical problem at the interface. A time-dependent problem is solved to obtain the temperature evolutions throughout all regions for 5 seconds.

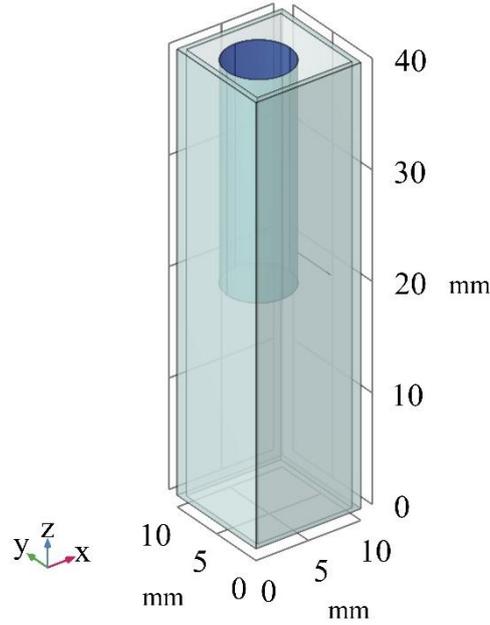

Figure 4. A schematic view of the geometry applied for the numerical simulations.

Table 2. Material properties considered for different parts of the numerical model.

| Region | Density ($\frac{kg}{m^3}$) | Thermal conductivity ($\frac{W}{m.K}$) | Heat capacity ($\frac{J}{kg.K}$) |
|---|---|---|---|
| cuvette | 2210 | 0.8 | 730 |
| cylindrical heater | 7000 | 50 | 420 |
| fluid | 1005 | ~ | 4250 |

Thermal conductivity calculation

In order to start the trial-and-error loop used in the third step of the beam displacement method, an initial guess for the thermal conductivity coefficient is assumed to be the thermal conductivity coefficient of the base fluid. A MATLAB code is utilized to solve the right-hand side of Eq. (5) after obtaining the temperature profiles from the simulations by the finite element method. As can be seen from the flowchart in Fig. 5, the experimental set-up yields the beam displacement on the left-hand side of Eq. (5), which is then, employed to calculate the thermal conductivity and continued the trial-and-error loop.

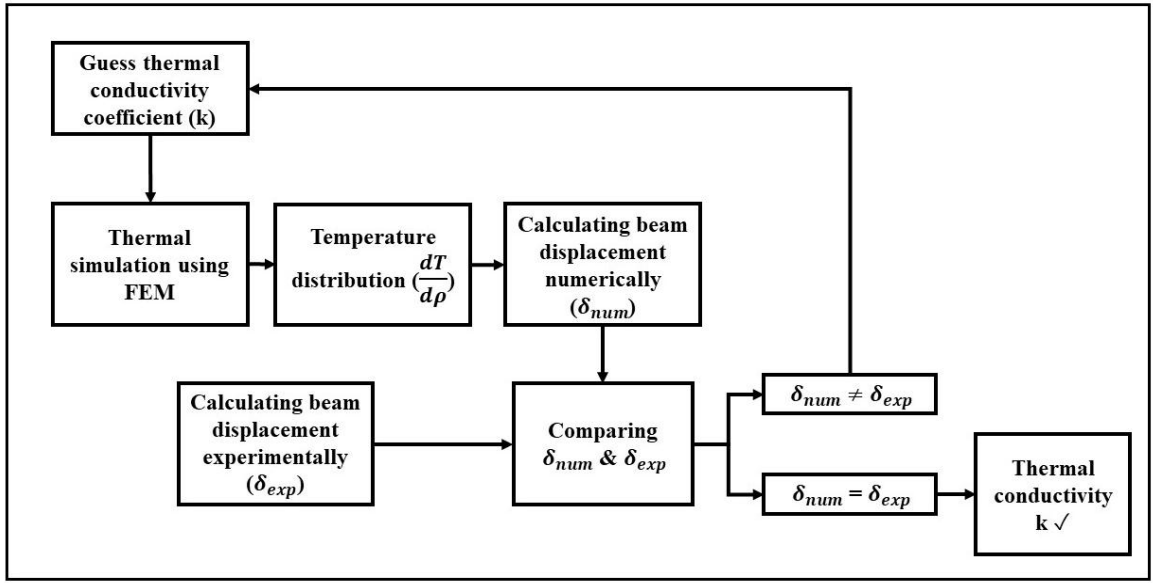

Figure 5. A schematic diagram outlining the procedures for the calculation of the thermal conductivity coefficient using beam displacement measurements and FE simulations.

Comparison of thermal conductivity measurements

A KD2 Pro apparatus was used at 25°C (Decagon Devices Inc., Pullman, WA, USA) to compare the thermal conductivity coefficients of the nanofluids with those obtained by the beam displacement method. The transient line heat source method underlies how the KD2 Pro analyzer operates. Before usage, the device was calibrated employing a standard sample of glycol. The measurements were performed after passing enough time for establishing the temperature equilibrium. A little amount of heat is applied to the needle by the KS-1 sensor, which aids in preventing free convection in liquid samples. Moreover, due to the sedimentation, the average thermal conductivity was measured three hours after preparation by repeating the tests five times.

### Electrical conductivity measurements

Electrical conductivity was estimated using the PCT-407 apparatus with a measuring range of 0-200 mS, and a 2% FS accuracy. Moreover, there is a nominal cell electrode in the PCT-407 device. The device was automatically calibrated using a calibration solution prior to conducting measurements. It should be noted that due to the sedimentation, the average electrical conductivity was determined six hours after preparation by repeating the tests five times.

# Result and discussion

## The stability results

### Observation Method

The observation method is used to investigate the stability of nanofluids by adding three surfactants including PVA, SDS, and CTAB. As demonstrated in Fig. 6, the suspension was unstable when PVA and SDS were used. Figs. 6(a) and 6(c) show the sedimentation observations of 0.025 vol.% $Al_{12}Mg_{17}$ nanofluids prepared immediately with two surfactants of PVA and SDS, and Figs. 6(b) and 6(d) illustrate the sedimentation observations after passing 30 minutes. The results indicate that SDS and PVA performed similarly in terms of improving stability for the 0.025 vol.% $Al_{12}Mg_{17}$ nanofluid and the nanoparticles agglomerated which started to sediment relatively shortly after the nanofluids were prepared. Contradictorily to PVA and SDS, CTAB demonstrated the required stability. Therefore, CTAB was used as a surfactant to prepare $Al_{12}Mg_{17}$ nanofluids at various concentrations.

Generally, surfactants prefer to be found at the interface between the nanoparticles and fluids because they create a sort of continuity between the two phases [39]. An interfacial layer will form around the nanoparticles as a result of the accurate amount of surfactants delivered into the nanofluid which is absorbed on the interface [40]. Studies have shown that adsorption is influenced by the characteristics of the solid substrate, the solvent, and the type of used surfactant [41]. It should be mentioned that SDS is an anionic surfactant [42], CTAB is a cationic surfactant [43], and PVP is a non-ionic polymer compound [44]. The compatibility of the surfactant depends on the surface charge of the $Al_{12}Mg_{17}$ nanoparticle, although PVA and SDS cannot be applied to the surface of $Al_{12}Mg_{17}$ nanoparticles, the results indicate that CTAB as a cationic surfactant can form a thin coating surrounding them.

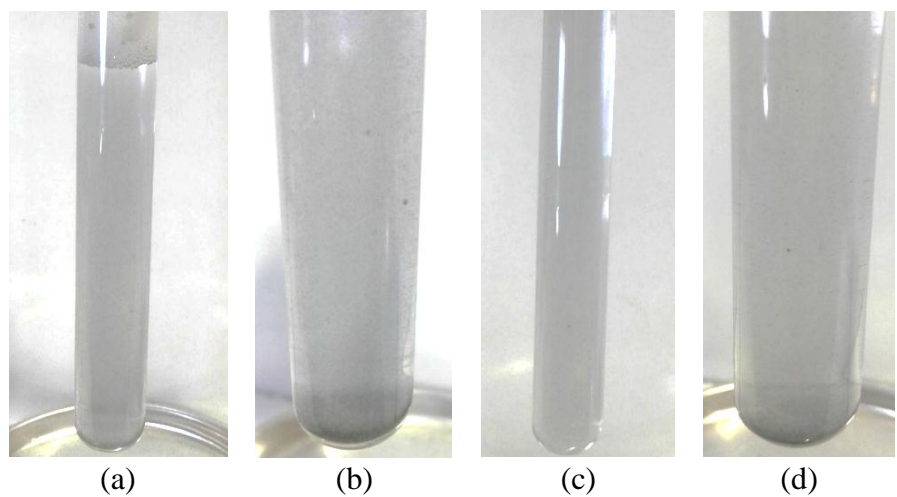

(a)  (b)  (c)  (d)

Figure 6. $Al_{12}Mg_{17}$ nanofluids with PVA (a) immediately after preparation, and (b) 30 minutes after preparation; and $Al_{12}Mg_{17}$ nanofluids with SDS (c) immediately after preparation, and (d) 30 minutes after preparation.

## Zeta potential analysis

The zeta potential of 0.025 vol.% $Al_{12}Mg_{17}$ nanofluid at various CTAB concentrations is shown in Fig. 7. As can be seen in Fig. 7, the stability of the $Al_{12}Mg_{17}$ nanofluid has increased by adding more CTAB. The zeta potential values of the nanofluids are in the ranges of 26.5 to 47.1 mv, in which the greatest zeta potential is associated with 0.1 vol.% CTAB, i.e. the nanofluid with a surfactant to nanoparticle volume ratio of 1:4. Generally, zeta potential measurement follows the electrophoretic behavior monitoring to assess the stability of nanofluids [45]. A high zeta potential value corresponds to strong repulsive forces, which imply great stability [32]. In nanofluids, a high surface charge density causes considerable repulsive forces [46]. It is due to the fact that a low surfactant concentration cannot completely cover nanoparticles, consequently, a charge imbalance develops, which causes nanoparticles to aggregate and precipitate [47].

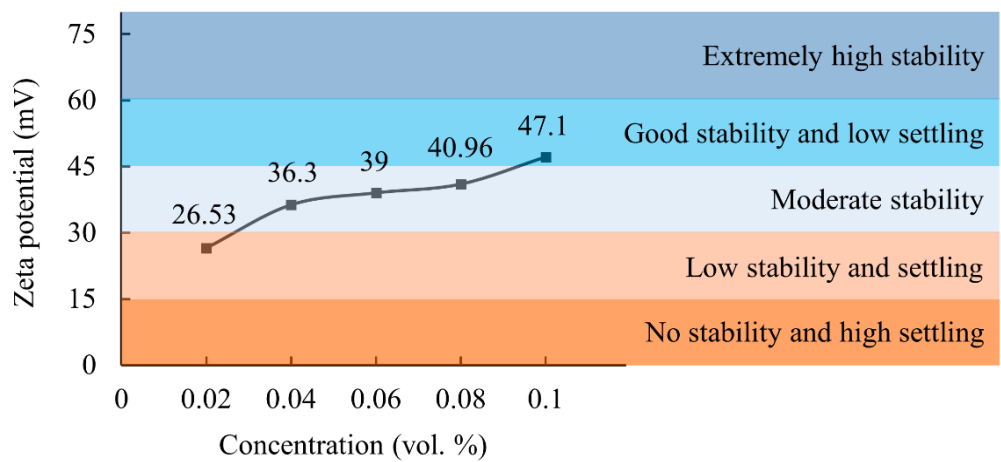

Figure 7. Zeta potential distribution of 0.025 vol.% $Al_{12}Mg_{17}$ nanofluids versus different concentrations of CTABs.

## Particle size distribution

Particle size distribution for 0.025 vol.% $Al_{12}Mg_{17}$ nanofluid at different ultrasonication periods is shown in Fig. 8. The TEM results illustrated that $Al_{12}Mg_{17}$ nanoparticles adhered to one another and formed large clusters, which should be broken apart using an ultrasonic wave. Fig. 8(a) represents the PSD for $Al_{12}Mg_{17}$ nanofluid after 15 minutes of ultrasonication, with a peak of 295 nm. Figs. 8(b) to 8(e) show the particle size distribution after 30, 45,60 and 75 minutes of ultrasonication, respectively, which indicate that the size of the nanoparticles remained noticeably unchanged. Moreover, Fig. 8(f) represents that the peak of PSD for $Al_{12}Mg_{17}$ nanoparticle decreases to 228 nm after 90 minutes of ultrasonication, while Figs. 8(g) and 8(h) show that when the ultrasonication period was increased to 105 min and 120 min, the peak of PSD decreased to 189 nm and 154 nm, respectively. As demonstrated in Fig. 8(h), the optimum ultrasonication period is about 120 minutes in which the peak of PSD reaches to 154 nm. It should be noted that after two hours, the peak of PSD grew with increasing ultrasonication period, reaching to a value of 276 nm for an ultrasonication period of 135 min, and finally shift to 700 nm for an ultrasonication period of 180 min (Figs. 8(i) to 8(l)).

Ultrasound sonication is a kind of vibration which provides the nanoparticle with a needed energy to release from the force for holding it in place. Therefore, extra energy is applied during sonication in the nanofluid to facilitate the movement of the nanoparticles. The nanoparticle cannot escape from the constriction force within the clusters if nanofluids don't obtain enough energy. On the other hand, the cluster will collide with other clusters more frequently if too much energy is expended for moving it. Therefore, each cluster would be more likely to entangle with and interact with other clusters, which would result in the formation of larger clusters [48]. Consequently, it is important to determine the optimum ultrasonication period for the nanofluids. This quantity depends on the type of nanoparticle, the type of ultrasound sonication, the ultrasound's power [40], and the ultrasound's pulse [49]. For example, it has been discovered that ultrasonic horn/probe devices are considerably more successful at dissolving the clusters rather than ultrasonic bath devices [50]. Continuous pulses, as another crucial factor affecting the dispersion of nanoparticles in fluid, can break up clusters into smaller pieces and the nanoparticles size distribution in the nanofluid will become more uniform. Discontinuous pulses, however, are unable to completely disperse the clusters, and some sizeable aggregates can still be found in the nanofluid [49]. As mentioned above, the ultrasound's power depends on the amount of energy which is needed for disintegrating the clusters to their constituent particles. Furthermore, for controlling the ultrasound's power it should be considered that receiving too much energy can cause nanoparticles to start re-agglomeration [51].

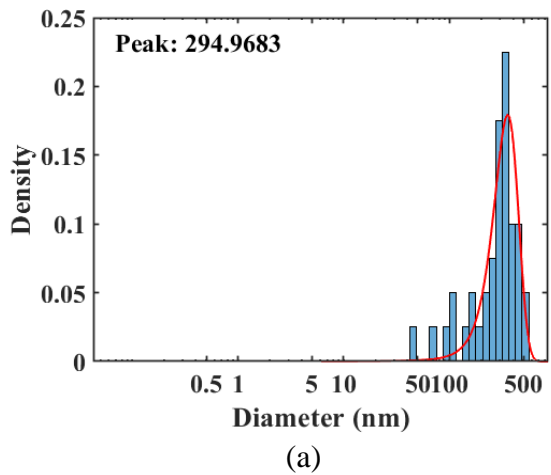
(a)

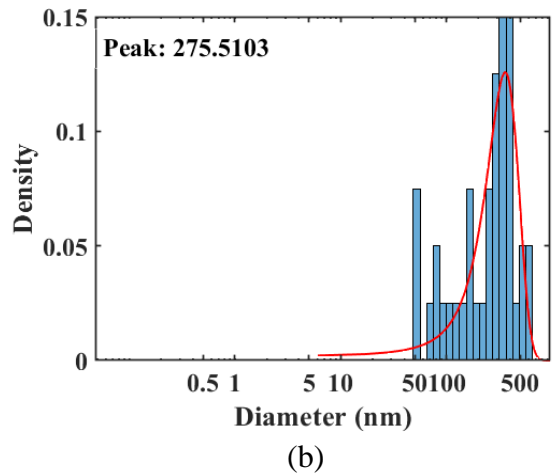
(b)

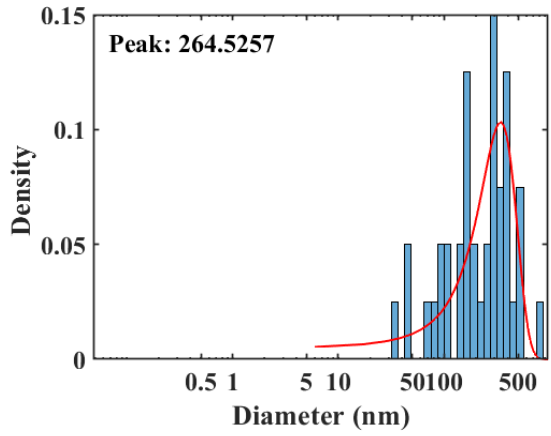
(c)
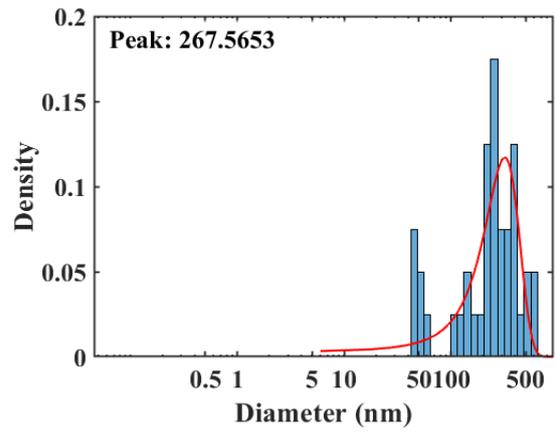
(d)
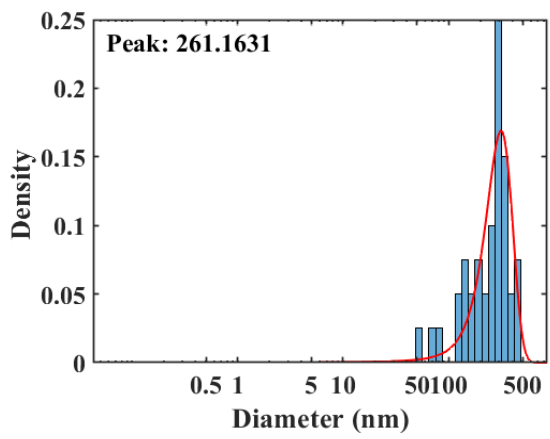
(e)
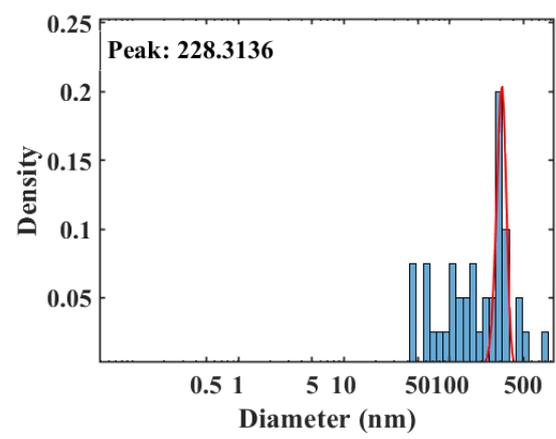
(f)
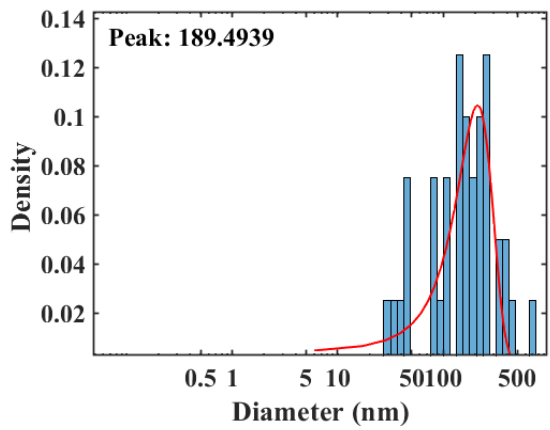
(g)
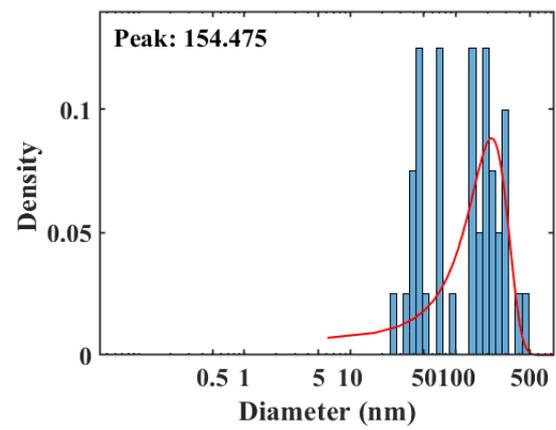
(h)

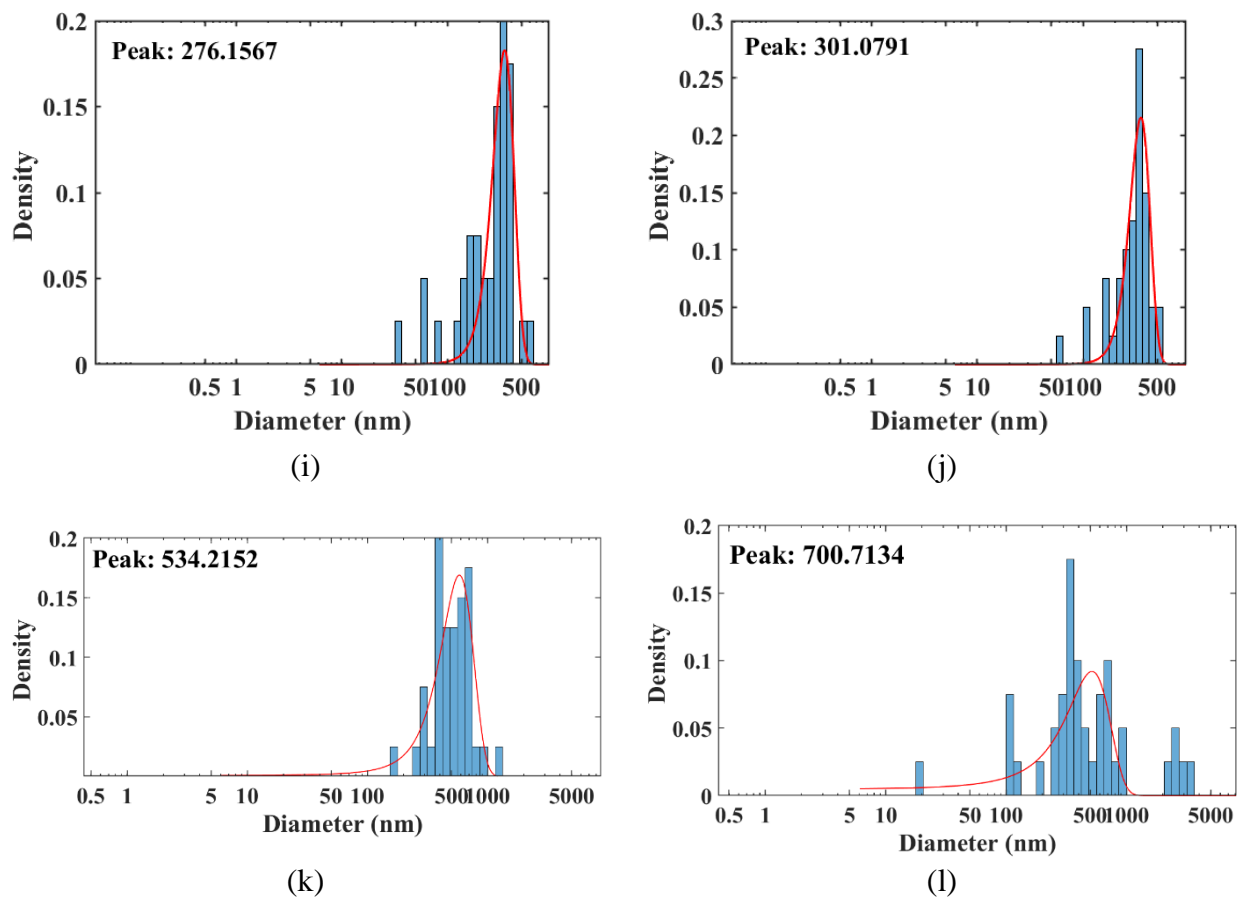

Figure 8. Particle size distribution of 0.025 vol.% $Al_{12}Mg_{17}$ nanofluid with 1 vol.% CTAB after ultrasonication periods of (a) 15 min, (b) 30 min, (c) 45 min, (d) 60 min, (e) 75 min, (f) 90 min, (g) 105 min, (h) 120 min, (i) 135 min, (j) 150 min, (k) 165 min, and (l) 180 min.

**Microstructural characterizations**

A drop of nanofluid was placed on a carbon grid for TEM scanning after the preparation of the nanofluid. In order to accurately analyze the nanoparticle size and its morphology in the nanofluid, and as well as to validate the particle size distribution, Fig. 9 illustrates a bright-field TEM micrograph of the 0.025 vol.% $Al_{12}Mg_{17}$ nanofluid at various ultrasonication periods. Figs. 9(a) and 9(b) show the range of distinct cluster sizes after 15 minutes of ultrasonication. It should be noted that although the clusters are spherical or nearly spherical, as seen in Fig. 9(a), their geometry is unknown, and it is estimated that their size is between 50 and 100 nm. Moreover, according to Fig. 9(b), the particle size of $Al_{12}Mg_{17}$ nanofluid exhibits non-uniformity with cluster sizes ranging from 150 to 500 nm. The TEM results after 120 minutes of ultrasonication are presented in Figs. 9(c) and 9(d) which represent that although the cluster sizes are not still uniform, these figures indicate that ultrasonication causes clusters to fragment and shrink in size.

Furthermore, the particle size distribution results demonstrated that they are not uniform, as seen in Fig. 8. It's worth mentioning that after 15 minutes of ultrasonication, the PSD measurements revealed a range of cluster sizes between 50 and 500 nm (Fig. 8(a)). Additionally, the particle size distribution after 120 minutes of ultrasonication, depicted in Fig. 8(g), demonstrates that the particle sizes decrease in comparison with the PSD results after 15 minutes of ultrasonication, while the particle size is not uniform. Besides, the ultrasonication process split the clusters and causes a drop in particle size. As shown in Fig. 9(a), the TEM results illustrate a range of cluster sizes between 50 and 300 nm for nanofluids after 15 minutes of ultrasonication. Moreover, as depicted in Fig. 9(c), the cluster size distribution after 120 minutes of ultrasonication has a prevailing size range of 110~190 nm which is corresponded to the PSD results in Fig. 8(h). According to Figs. 9(c) and 9(d), the TEM results present a wide variety of sizes for nanoparticles, too. In this way, the TEM results represented in Fig. 9 provide support for the PSD tests reported in Fig. 8, so validating them.

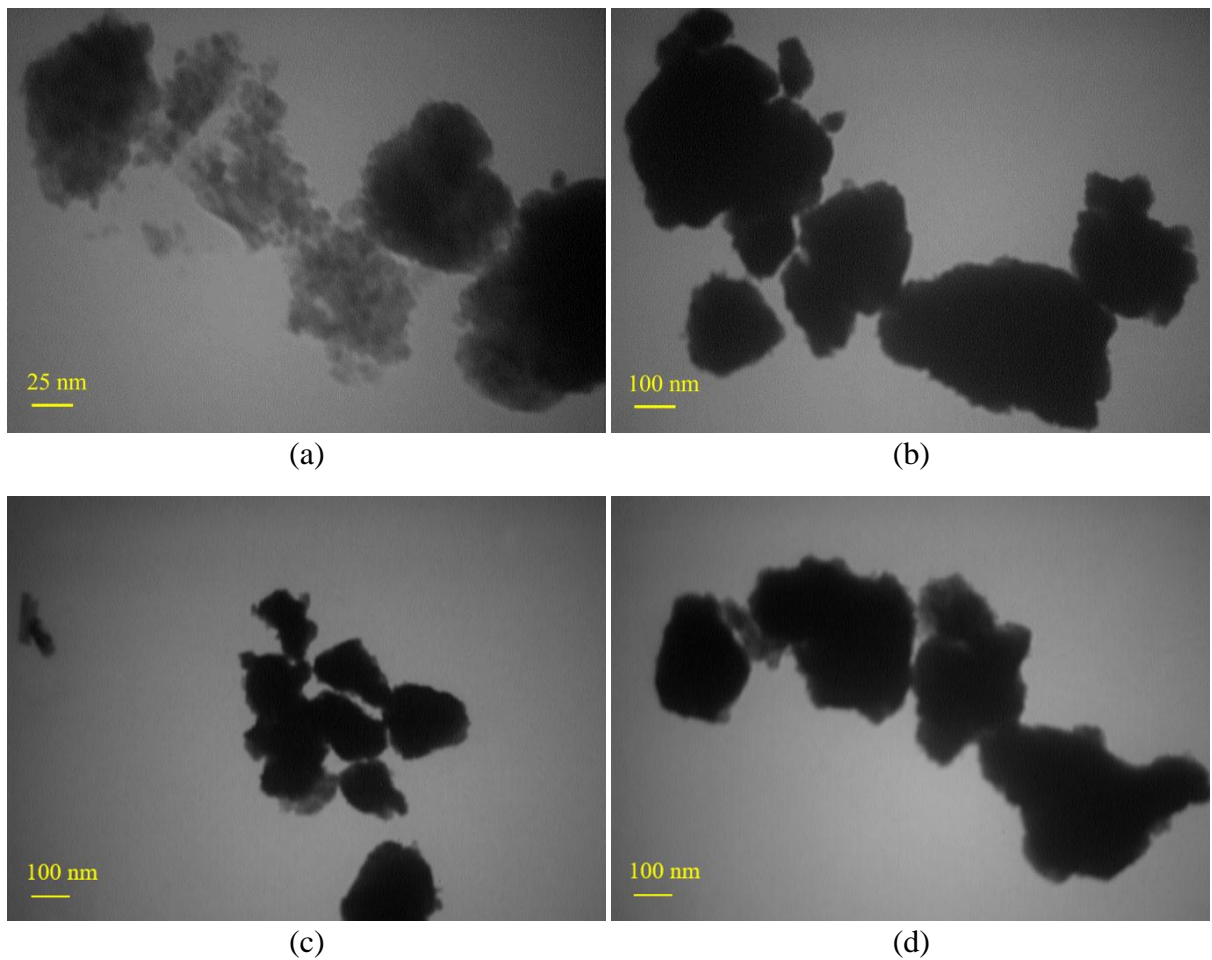

Figure 9. TEM results for the 0.025 vol.% $Al_{12}Mg_{17}$ nanoparticles in DI water. Distribution of nanoparticles after 15 minutes of ultrasonication: (a) a view of smaller-sized nanoparticles, and (b) a view of larger-sized nanoparticles; and after 120 minutes of ultrasonication: (c) a view of smaller-sized nanoparticles, (d) a view of larger-sized nanoparticles.

### Thermal conductivity coefficient

As previously described, the beam displacement approach is used to determine the thermal conductivity of the $Al_{12}Mg_{17}$ nanofluid in three steps. As the results of the first step, the deviation of the beam due to the thermal shock is shown in Fig. 10. This figure shows the beam displacement measurements for various concentrations of $Al_{12}Mg_{17}$ nanoparticles versus the number of sampling frames. As shown in Fig. 10, beam displacements of 13.37, 13.53, 13.7, and 13.8 $\mu m$ were measured for 0.0125, 0.025, 0.0375, and 0.05 vol.% $Al_{12}Mg_{17}$ nanofluids, respectively. Beam displacements were calculated as the difference between the average value of the beam spot position in the steady state and the displacement peak.

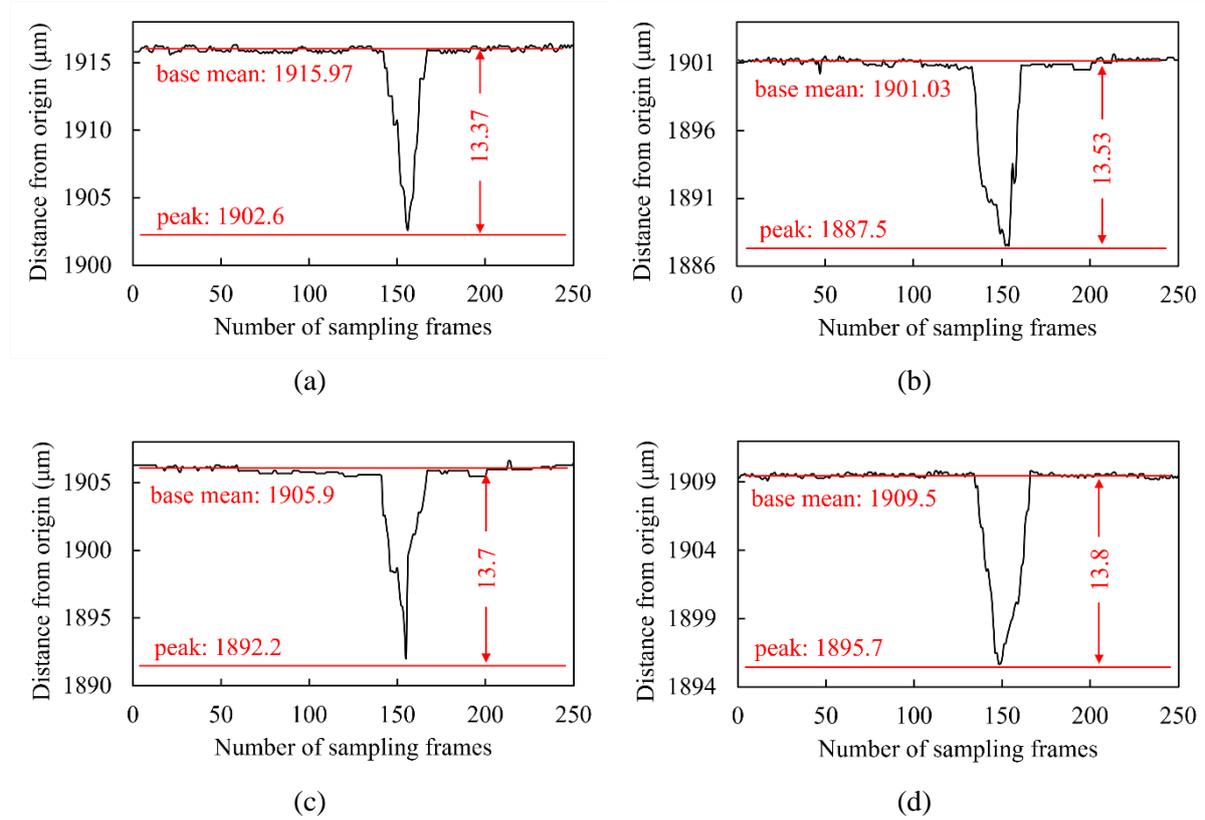

Figure 10. Beam displacement results for different concentrations of $Al_{12}Mg_{17}$ nanofluid at (a) 0.0125, (b) 0.025, (c) 0.0375, and (d) 0.05 vol.%.

As the outcome of the second step, temperature variations through a line corresponding to the beam path are numerically simulated. Fig. 11 shows the thermal variations through the domain for the thermal conductivity coefficient of 0.8 $W/(m.K)$ after 5 s. Figs. 11(a) and 11(b) show the isothermal contour and the temperature variation on a cut plane across the center of the heater (at y=5 mm). This figure indicates that after 5 seconds of heating, there is a temperature difference of approximately 70 degrees close to the heater, which is the reason for the beam deviation.

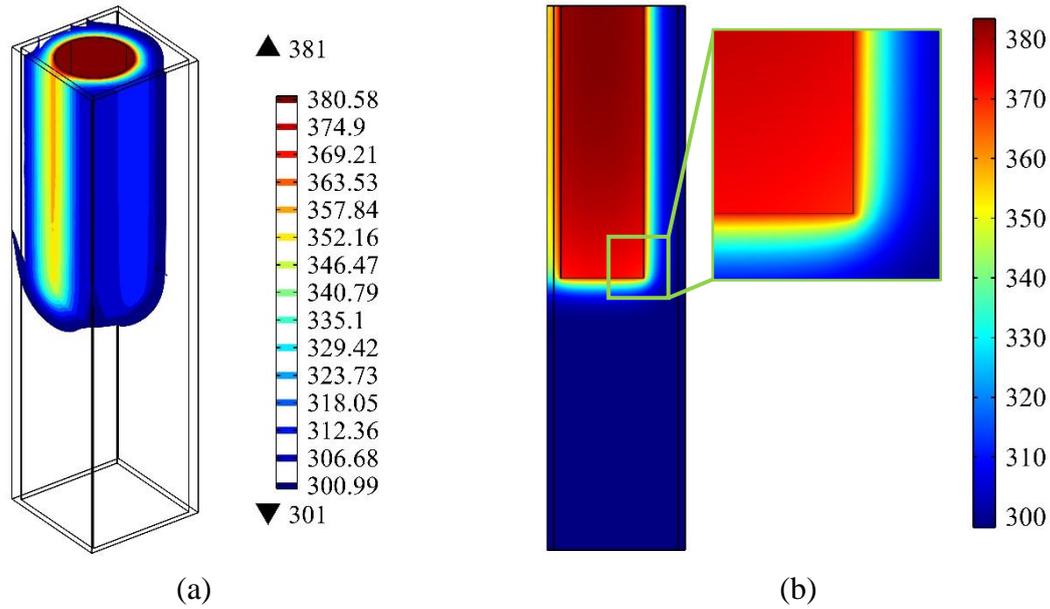

(a)                                    (b)

Figure 11. (a) Isothermal contour of the domain, and (b) temperature distribution on a cut plane parallel to xz-plane at y=5 mm for the nanofluid's thermal conductivity coefficient of $0.8\ W/(m.K)$ after 5 s.

Following the third step, after the comparison of the experimental results with the numerical simulations, the thermal conductivity of the nanofluid was obtained. Therefore, the results of the beam displacement method demonstrated that the thermal conductivity of $Al_{12}Mg_{17}$ nanofluid is $0.61\ W/(m.K)$ at a concentration of 0.0125 vol.% and at concentrations of 0.025, 0.0375, and 0.05 vol.% increases to 0.66, 0.73, and $0.8\ W/(m.K)$, respectively.

The temperature variations on the beam path line (Fig. 12(a)) for different values of the thermal conductivity coefficient of the nanofluid are depicted in Fig. 12(b). From this figure, it can be concluded that for a 0.05 vol.% $Al_{12}Mg_{17}$ nanofluid with a thermal conductivity coefficient of $0.8\ W/(m.K)$, the temperature is about 364 K at a distance of 100 microns from the heater after 5 seconds of heating, while it is approximately 368 K for a 0.0125 vol.% $Al_{12}Mg_{17}$ nanofluid with a thermal conductivity value of $0.61\ W/(m.K)$. Thus, heat transfer is improved by increasing the concentration of nanoparticles. Fig. 12(d) shows the temperature contour of the nanofluid with a thermal conductivity coefficient of $0.8\ W/(m.K)$ on a cut plane parallel to xy-plane that crosses the beam path line (Fig. 12(c)). Given that fluids have a larger refractive index at lower temperatures, the nanofluid with the thermal conductivity coefficient of $0.8\ W/(m.K)$ has the lowest temperature changes among all and therefore has the largest displacement.

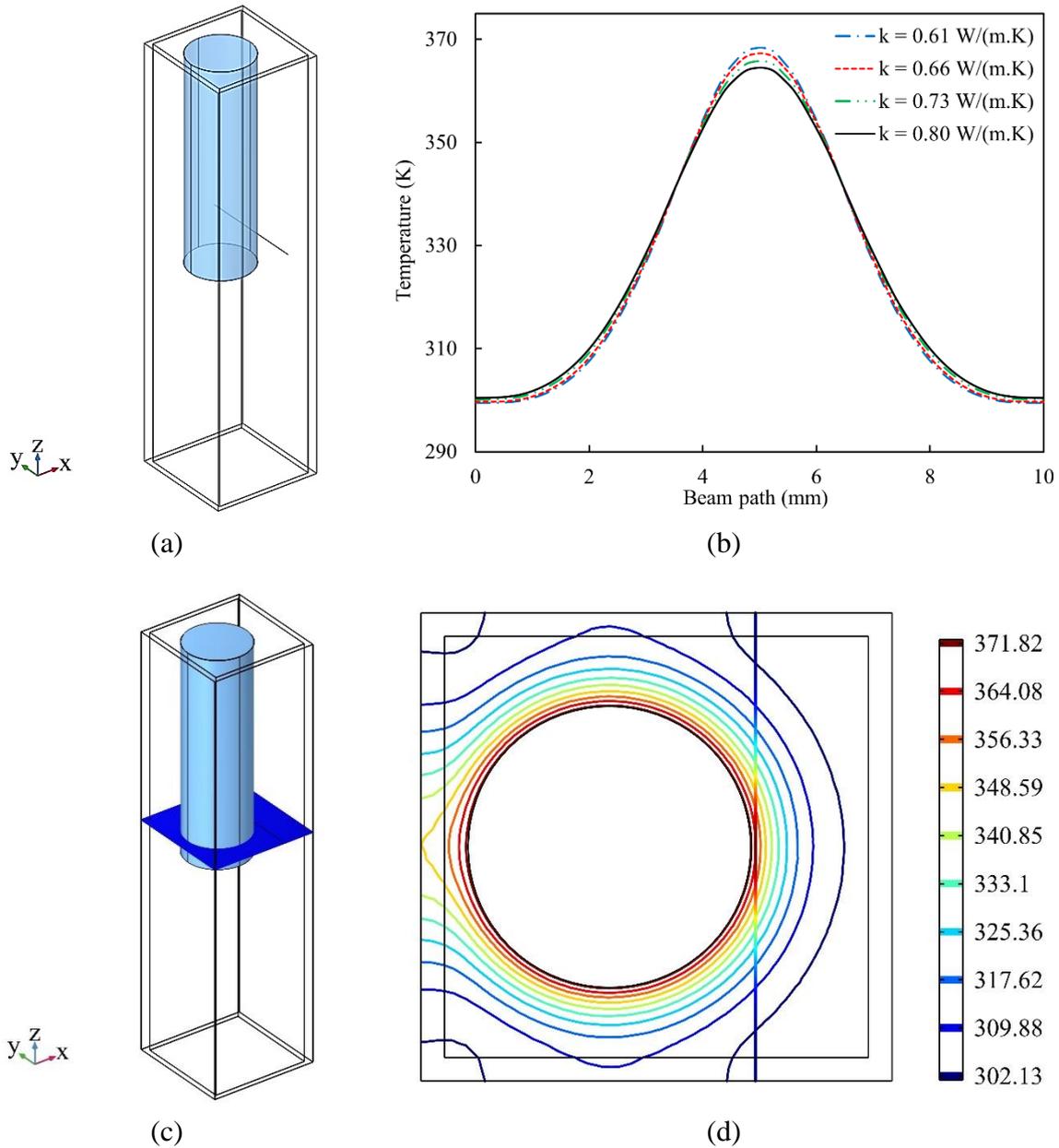

Figure 12. (a) A view of the beam path line (at $x = 7.1\ mm, y = 0{\sim}10\ mm, z = 22\ mm$), and (b) temperature variations on the beam path line after 5 s for different values of the nanofluid's thermal conductivity coefficient. (c) A view of the cut plane parallel to xy-plane at z=22 mm, and (d) temperature contour on the cut plane and the beam path line for $k = 0.8\ W/(m.K)$ after 5 s.

In addition, the thermal conductivity measurements by the KD2 Pro apparatus were compared with the results obtained using the beam displacement approach. Fig. 13 shows the thermal conductivity of the nanofluids measured using the KD2 Pro instrument at $25°C$ and the beam displacement technique at various concentrations of $Al_{12}Mg_{17}$ nanoparticles. According to the KD2 Pro results, the thermal conductivity of the 0.0125 vol.% $Al_{12}Mg_{17}$ nanofluid is $0.633\ W/(m.K)$, which rises to 0.71, 0.78, and 0.81 $W/(m.K)$ by increasing the concentration of $Al_{12}Mg_{17}$ nanoparticles to 0.025, 0.0375, and 0.05 vol.%, respectively.

Based on the measured thermal conductivity values, the enhancements in the thermal conductivities were estimated relative to the base fluid as shown in Fig. 13(b). The findings demonstrate that, in comparison to DI water with a thermal conductivity coefficient of 0.579 $W/(m.K)$, the thermal conductivity is higher when nanoparticles are present. With $Al_{12}Mg_{17}$ nanoparticles distributed in DI water at a concentration of 0.05 vol.%, the highest overall improvement in thermal conductivity of almost 40% and 38% was observed using KD2 Pro measurement and beam displacement method, respectively. On the other hand, the 0.0125 vol.% $Al_{12}Mg_{17}$ nanofluid exhibits the lowest magnitude of the thermal conductivity enhancement with a value of approximately 9% and 5% using KD2 Pro measurement and beam displacement method, respectively.

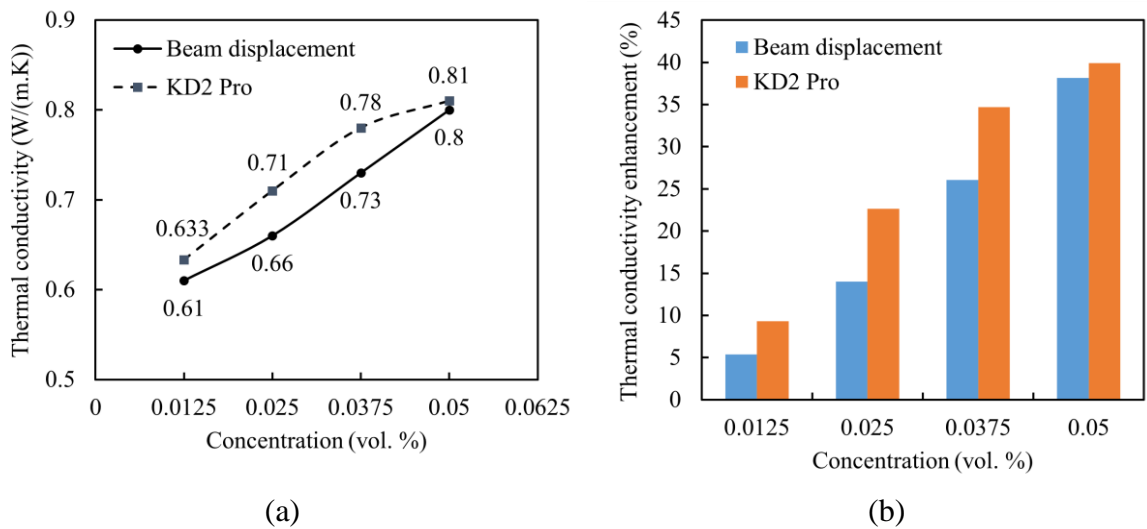

(a)                                                            (b)

Figure 13. (a) Dependence of thermal conductivity of $Al_{12}Mg_{17}$ nanofluids on the concentration of nanoparticles with 0.1 vol.% CTAB. (b) Thermal conductivity enhancement of the $Al_{12}Mg_{17}$ nanofluids in comparison with DI water.

Different factors of the Brownian motion, the cluster size, the aggregation of nanoparticles, the formation of a layer of fluid molecules close to the surfaces of the nanoparticles, the formation of nanoparticle complexes, and the collisions between clusters may all contribute to the increased heat transfer in nanofluids containing alloy nanoparticles [52]. It should be noted that thermal conductivity also depends on the crystal size, volume fraction of nanoparticles, and thermal characteristics of the solid suspension [53].

Table. 3 shows a comparison between the thermal conductivity enhancements of the nanofluid examined in the present study and the results of two other studies. In the study of Paul et al. [20], they noticed a 16% increase in thermal conductivity after dispersing 0.1 vol.% of $Al_{95}Zn_{05}$ nanoparticles in ethylene glycol. They asserted that the increased rate of heat transfer in nanofluids is attributed to the large specific surface area of the nanoparticles, the particle shape factor, liquid layering at the solid-liquid interface, clustering/aggregation, and the Brownian motion. Furthermore, the research conducted by Karthik et al. [24] indicated that the addition of 0.1 vol.% $Ni_{65}Al_{35}$ nanoparticles boosted the thermal conductivity of nanofluid by 28%. This study showed that the $Ni_{65}Al_{35}$ intermetallic surface composition and the bulk stoichiometry had little effect on the increase of the thermal conductivity of nanofluids containing $Ni_{65}Al_{35}$ nanoparticles. Therefore, in the present study, $Al_{12}Mg_{17}$ nanoparticles distributed in DI water outperform $Ni_{65}Al_{35}$ and $Al_{95}Zn_{05}$ nanoparticles dispersed in water and ethylene glycol, respectively.

Table 3. A comparison among the thermal conductivity enhancements reported by different references and the current research.

| Study | Base fluid | Nanoparticle | Concentration | Thermal conductivity enhancement |
|---|---|---|---|---|
| Paul et al. [20] | ethylene glycol | $Al_{95}Zn_{05}$ | 0.1 vol.% | 16% |
| Karthik et al. [24] | water | $Ni_{65}Al_{35}$ | 0.1 vol.% | 28% |
| Present study | water | $Al_{12}Mg_{17}$ | 0.05 vol.% | 40% |

**Electrical conductivity enhancement**

The results of the electrical conductivity measurements for the $Al_{12}Mg_{17}$ nanofluid at different concentrations of nanoparticles are shown in Fig. 14. In the absence of $Al_{12}Mg_{17}$ nanoparticles, a base fluid containing 0.1 vol.% CTAB has an electrical conductivity of 86 $\mu S/cm$. As can be clearly seen in Fig. 14, the electrical conductivity of $Al_{12}Mg_{17}$ nanofluid increases linearly from 155 to 188 $\mu S/cm$ by enhancing the volume fraction from 0.0125 to 0.05 vol.%. The enhancements in the electrical conductivities of nanofluids were calculated relative to the base fluid containing 0.1 vol.% CTAB. The maximum enhancement in electrical conductivity of about 116% was observed for the 0.05 vol.%. $Al_{12}Mg_{17}$ nanofluid. The electrical conductivity of a nanofluid is associated with the ability of the charged ions inside the nanofluid to transport electrons. This might be due to the possibility of an electrical double-layer formation on the surface of the dispersed nanoparticles [54]. The primary cause of the increase in electrical conductivity is the creation of surface charges caused by the polarization of nanoparticles when dispersed in the polar water. The dispersion of the nanoparticles alters the dielectric constant and density of the base fluid. Therefore, it seems reasonable that an increase in the electrical conductivity would follow a rise in the concentration of $Al_{12}Mg_{17}$ nanoparticles.

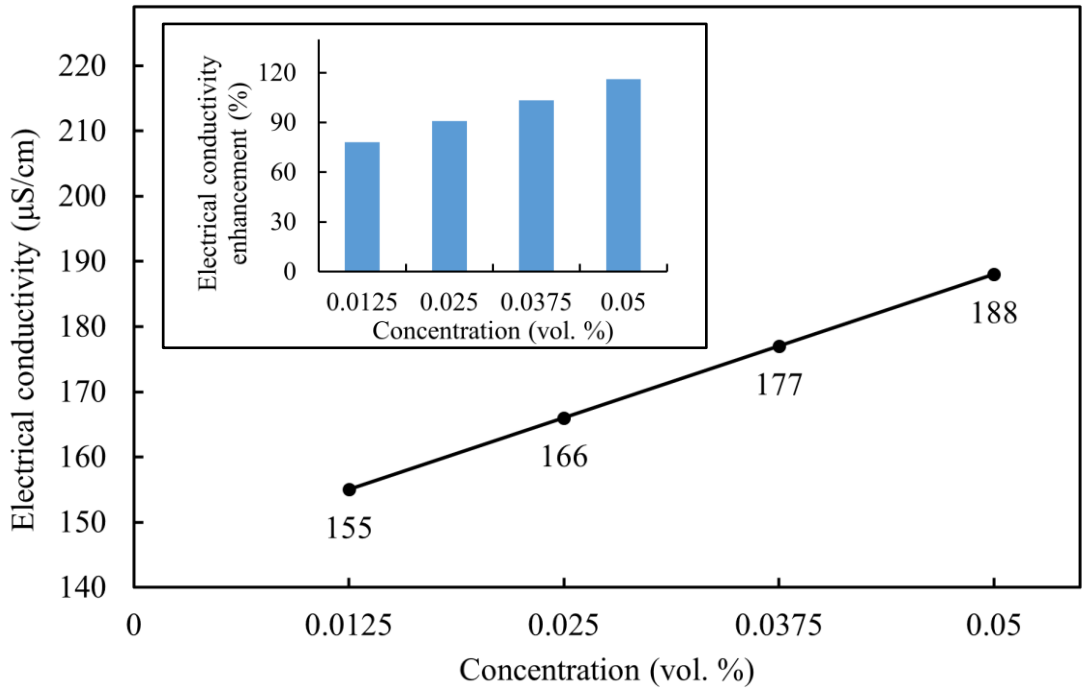

Figure 14. The effective electrical conductivity of $Al_{12}Mg_{17}$ nanofluid at various concentrations of nanoparticles with 0.1 vol.% CTAB at 25°C.

## Conclusion

In this study, experiments were conducted to determine the stability, optimum ultrasonication period, thermal conductivity, and electrical conductivity of the $Al_{12}Mg_{17}$ nanofluid. The findings of this research provide engineers and scientists with our perspective on the potential of nanofluids as coolants in heat transfer devices such as solar collectors, automobile radiators, and electronic cooling. It should be mentioned that previous researchers have paid less attention to how the particle size distribution evolves with the ultrasonication period in order to establish the optimum ultrasonication period.

The particle size distribution of $Al_{12}Mg_{17}$ nanofluid was determined using the dynamic light scattering method, and then, confirmed by TEM micrograph imaging. Visual observation and zeta potential measurement were employed to assess the stability of the nanofluids. A brand-new method was also introduced for the measurement of the thermal conductivity coefficient using the experimental evaluation of the beam displacement and numerical analysis. One advantage of the optical methods for thermal characterization is the ability to quickly collect the results before the appearance of bulk motion in the nanofluid which occurs due to thermal variations. KD2 Pro apparatus was used to validate the thermal conductivity data obtained from the new beam displacement method. In addition, PCT 407 was exploited to measure electrical conductivity. In order to review the outcomes, the results obtained from the current research are summarized in the following:

- PVA and SDS were not suitable surfactants for stabilizing $Al_{12}Mg_{17}$ nanoparticles within DI water,

- CTAB was chosen as the most appropriate surfactant to prepare stable nanofluids with a concentration range of 0.0125 to 0.05 vol.% by utilizing the stirrer and ultrasonication processes,
- The particle size distribution showed an optimum ultrasonication period of 2 hours with a peak of 154 nm, and the TEM results confirmed the PSDs results obtained from DLS experiments,
- A maximum zeta potential of 47.1 mv was measured for a 0.025 vol.% of $Al_{12}Mg_{17}$ nanofluid containing 0.1 vol.% CTAB,
- The intermetallic nanoparticles similar to other nanoparticles improved the thermal conductivity of nanofluid and the highest thermal conductivity enhancement of 40% was observed at a concentration of 0.05 vol.% $Al_{12}Mg_{17}$ nanoparticles,
- The highest electrical conductivity of 188 $\mu S/cm$ was observed at a concentration of 0.05 vol.% $Al_{12}Mg_{17}$ nanoparticles.

Therefore, it can be concluded that $Al_{12}Mg_{17}$ nanofluid is a special nanofluid with the potential to be useful for applications in heat transfer. In this regard, a cutting-edge technique for determining the thermal conductivity coefficient is the novel beam displacement approach presented in the current study.

## Acknowledgement
The authors are so grateful from Professor Ali Shokuhfar who passed away incredibly these days. We all learned a lot from his high knowledge, great personality, and kind behavior.

## Author contributions

Soroush Javadipour designed and performed the experiments and prepared the initial version of the manuscript with data analysis (signal processing with MATLAB and image processing

**Competing interests**

The authors declare no competing interests.